# Optimal Entropic Uncertainty Relation for Successive Measurements in Quantum Information Theory

M. D. Srinivas

*Centre for Policy Studies*
*27, Rajasekaran Street, Mylapore, Chennai 600004, India*
e-mail: policy@vsnl.com
(August 1, 2002)

We derive an optimal bound on the sum of entropic uncertainties of two or more observables when they are sequentially measured on the same ensemble of systems. This optimal bound is shown to be greater than or equal to the bounds derived in literature on the sum of entropic uncertainties of two observables which are measured on distinct but identically prepared ensembles of systems. In the case of a two-dimensional Hilbert Space, the optimal bound for successive measurements of two spin components, is seen to be strictly greater than the optimal bound for the case when they are measured on distinct ensembles, except when the spin components are mutually parallel or perpendicular.

# 1. Uncertainty relations for Distinct and Successive Measurements

Heisenberg, in his celebrated paper on uncertainty relations [1], written seventy-five years ago, seems to have interpreted these relations as expressing the influence of measurement of one observable on the uncertainty in the outcome of a succeeding measurement of another observable performed on the same system. Heisenberg states:

> At the instant when position is determined – therefore, at the moment when the photon is scattered by the electron – the electron undergoes a discontinuous change in momentum. This change is the greater the smaller the wavelength of the light employed – that is, the more exact the determination of the position. At the instant at which the position of the electron is known, its momentum therefore can be known up to magnitudes, which correspond to that discontinuous change. Thus, the more precisely the position is determined, the less precisely the momentum is known, and conversely.

However, the standard mathematical formulation of the uncertainty relations, as first derived by Robertson in 1929 [2] from the first principles of quantum theory, does not in any way support the above interpretation. If $A$, $B$ are two observables and $\rho$ is the density operator characterizing the state of the system, then the standard formulation of the uncertainty relation is given by the inequality

$$(\Delta^\rho A)^2 (\Delta^\rho B)^2 \geq \tfrac{1}{4} | \operatorname{Tr}(\rho[A, B])|^2 , \tag{1}$$

where, the variances $(\Delta^\rho A)^2$ and $(\Delta^\rho B)^2$ are given by

$$(\Delta^\rho A)^2 = \operatorname{Tr}(\rho A^2) - [\operatorname{Tr}(\rho A)]^2 \tag{2a}$$

$$(\Delta^\rho B)^2 = \operatorname{Tr}(\rho B^2) - [\operatorname{Tr}(\rho B)]^2 . \tag{2b}$$

Since the variance $(\Delta^\rho A)^2$ of an observable $A$ is nonnegative and vanishes only when the state $\rho$ is a mixture of eigenstates of $A$ associated with the same eigenvalue, it is a reasonable measure of the uncertainty in the outcome of an $A$-measurement carried out on an ensemble of systems in state $\rho$. However, $(\Delta^\rho A)^2$ and $(\Delta^\rho B)^2$ as given by eqs. (2a), (2b) refer to uncertainties in the outcomes of $A$ and $B$ when they are measured on *distinct* though *identically prepared* ensembles of systems in state $\rho$. Thus the inequality (1) refers to a situation where the observables $A$, $B$ are measured on distinct ensembles of systems and may hence be referred to as *Uncertainty Relation for Distinct Measurements*.

The current understanding of uncertainty relations (1) is succinctly summarised in the recent treatise on quantum computation and quantum information by Nielsen and Chuang as follows [3] (we have altered the symbols appearing in the quotation so as to correspond to eq. (1) above; emphasis as in the original):

> You should be vary of a common misconception about the uncertainty principle, that measuring an observable $A$ to some "accuracy" $\Delta^\rho A$ causes the value of B to be "disturbed" by an amount $\Delta^\rho B$ in such a way that some sort of inequality similar to (1) is satisfied. While it is true that measurements in quantum mechanics cause disturbance to the system being measured, this is most emphatically *not* the content of the uncertainty principle.

The correct interpretation of the uncertainty principle is that if we prepare a large number of quantum systems in identical states, $\rho$, and then perform measurements of A on some of those systems and of B on others, then the standard deviation $\Delta^\rho A$ of the A results times the standard deviation $\Delta^\rho B$ of the results for B will satisfy the inequality (1).

In fact, the uncertainty relation (1) essentially expresses the limitations imposed by quantum theory on the preparation of ensembles of systems in identical states, as has been clearly explained in the recent monograph on the conceptual foundations of quantum theory by Home [4]:

> To test uncertainty relation …(1), we require a repeatable state preparation procedure leading to an ensemble of identically prepared particles, all corresponding to the state being studied. Then on each such system we must measure one or the other of the two dynamical variables (either *A* or *B*). The statistical distributions of the measured results of *A* and *B* are characterised by variances satisfying …(1). The significance of the uncertainty principle is stated as follows: It is impossible to prepare an ensemble of identical particles, all in the same state, such that the product of …[variances] of any two noncommuting dynamical variables has a value less than the lower bound given by relation …(1).

In order to explore the influence of the measurement of one observable on the uncertainties in the outcomes of another, we have to formulate an *Uncertainty Relation for Successive Measurements*, which is different from (1). For this purpose, we will have to consider the uncertainties in the outcomes of the *A* and *B*-measurements, when they are performed *sequentially* on the *same* ensemble. If *A* and *B* are observables with purely discrete spectra with spectral resolution

$$A = \sum_i a_i P^A(a_i) \qquad (3a)$$

$$B = \sum_j b_j P^B(b_j), \qquad (3b)$$

then the joint probability $\Pr^\rho_{A,B}(a_i, b_j)$ that the outcomes $a_i, b_j$ are realised in a sequential measurement of the observables *A, B,* is given by the well-known Wigner formula [5,6]:

$$\Pr^\rho_{A,B}(a_i, b_j) = \text{Tr}\,[P^B(b_j)P^A(a_i)\,\rho\,P^A(a_i)P^B(b_j)]. \qquad (4)$$

In the above, and in what follows, we adopt the Heisenberg picture of time evolution, where the observables carry the entire burden of time evolution in the absence of measurements. An uncertainty relation for successive measurements can easily be derived [7] by calculating the variances $\Delta^\rho_{A,B}(A)$ and $\Delta^\rho_{A,B}(B)$, which correspond to the uncertainties in the outcomes of these observables when they are measured sequentially, by employing the joint probability (4):

$$\Delta^\rho_{A,B}(A)^2 \, \Delta^\rho_{A,B}(B)^2 \geq |\operatorname{Tr}[\rho A \mathcal{E}(B)] - \operatorname{Tr}[\rho A]\operatorname{Tr}[\rho \mathcal{E}(B)]|^2, \qquad (5)$$

where

$$\mathcal{E}(B) = \sum_i P^A(a_i) \, B \, P^A(a_i). \qquad (6)$$

In a seminal paper, Deutsch [8] pointed out that, except in the case of canonically conjugate variables, the inequality (1) does not adequately express the quantum uncertainty principle. The RHS of (1) crucially depends on the state $\rho$ of the system except in the case of canonically conjugate observables. In order to get a nontrivial lower bound on the uncertainty in the outcome of $B$ given the uncertainty in the outcome of $A$, we need to take the infimum of RHS of (1) over all states $\rho$, and this invariably vanishes whenever $A$ or $B$ has a single discrete eigenvalue. Therefore the inequality (1) does not give any nontrivial lower bound on the uncertainties in most physical situations involving angular momenta, spin or finite-level systems. The variance form of the uncertainty relation for successive measurements (5) is even more ineffective, for the infimum of RHS of (5) identically vanishes as the operators $A$ and $\mathcal{E}(B)$ always commute.

Deutsch argued that, in order to have a meaningful formulation of the uncertainty principle, we should have a nontrivial lower bound on the product/sum of the uncertainties of two observables, which, unlike as in (1), is independent of the state and vanishes essentially only when the two observables have a common eigenvector. He also showed that it is possible to formulate such an uncertainty relation by making use of the information-theoretic entropy, instead of variance, as the measure of uncertainty.

## 2. Entropic Uncertainty Relations for Distinct Measurements

For a classical discrete random variable $X$ which takes on $n$ different values with associated probabilities $\{p_i\}$, the information-theoretic entropy given by

$$S(X) = -\sum_i p_i \log(p_i), \qquad (7)$$

is a good measure of the uncertainty, or spread in the probability distribution. In (7) we assume that $0 \log(0) = 0$ always. Since, $0 \leq p_i \leq 1$, and $\sum p_i = 1$, we have

$$0 \leq S(X) \leq \log(n). \qquad (8)$$

The lower bound in (8) is attained when all the $p_i$ vanish except for some $p_k$, $1 \leq k \leq n$; the upper bound is attained when all the $p_i$ are equal to $1/n$.

The above definition can be extended to cases where the number of outcomes is not finite and also to the case when $X$ is a continuous random quantity with the probability density $p(x)$. In the later case the information-theoretic entropy is usually defined as follows [9]:

$$S(X) = -\int p(x) \log p(x)\, dx. \tag{9}$$

Apart from the fact that $S(X)$ as defined above is not non-negative, it is also not physically meaningful as it has inadmissible physical dimensions. While the discrete probabilities $\{p_i\}$ and the corresponding entropy (7) are mere dimensionless numbers, the probability density $p(x)$ has dimension $(1/D)$, where $D$ is the physical dimension of the random quantity $X$, and hence $S(X)$ as defined by eq. (9) has inadmissible physical dimension $\log(D)$. It has therefore been suggested [10] that for continuous variables we should employ, instead of $S(X)$, its exponential $E(X)$, given by

$$E(X) = \mathrm{Exp}\left[-\int p(x) \log p(x)\, dx\right]. \tag{10}$$

The exponential entropy (10) has a physically meaningful dimension $D$, the same as the random quantity $X$. It is a monotonic function of $S(X)$ and is also nonnegative.

It may be of interest to note that the first entropic uncertainty relation in quantum theory was derived much prior to the work of Deutsch, and it was formulated for continuous observables, position and momentum. An entropic uncertainty relation for position and momentum, was conjectured by Everett in his famous thesis [11] in 1957 and by Hirschman [12] in the same year. It was proved in 1975 by Beckner [13] and Bialynicki-Birula and Mycielski [14]. If $\psi(x)$ is the wave function of a particle in one dimension, and $\varphi(p)$ the corresponding momentum space wave function, then the entropic uncertainty relation obtained by these authors may be expressed in the form

$$-\int |\psi(x)|^2 \log |\psi(x)|^2\, dx - \int |\varphi(p)|^2 \log |\varphi(p)|^2\, dp \geq \log(\pi \hbar e). \tag{11}$$

Clearly, both the LHS and RHS in eq. (11) have inadmissible physical dimensions. However, this can be corrected by re-expressing the above relation in terms of the exponential entropies as follows:

$$E^\rho(Q)\, E^\rho(P) \geq \pi \hbar e. \tag{12}$$

Eq. (12) is the appropriate entropic uncertainty relation for position and momentum, in fact for any pair of canonically conjugate observables, and it can also be shown to be stronger than the well-known variance form for the uncertainty relation for such observables.

We now turn to the recent work on the entropic uncertainty relations for observables with purely discrete spectra following the work of Deutsch. Let us consider two observables $A$, $B$, with purely discrete spectra and spectral resolution as given by (3a), (3b). Then the entropic uncertainties of $A$ and $B$ as measured in distinct but identically prepared ensembles in state $\rho$, are given by

$$S^\rho(A) = - \sum_i \text{Tr}\,[\rho P^A(a_i)]\,\log \text{Tr}\,[\rho P^A(a_i)] \tag{13a}$$

$$S^\rho(B) = - \sum_j \text{Tr}\,[\rho P^B(b_j)]\,\log \text{Tr}\,[\rho P^B(b_j)]\,. \tag{13b}$$

It can easily be seen that the entropic uncertainty $S^\rho(A)$ (as well as $S^\rho(B)$) is non-zero and vanishes only when the state $\rho$ is a mixture of the eigenstates of $A$ (correspondingly $B$), all associated with the same eigenvalue. Now the optimum uncertainty relation for distinct measurements is of the form

$$S^\rho(A) + S^\rho(B) \geq \Lambda_D(A, B), \tag{14}$$

where

$$\Lambda_D(A, B) = \underset{\rho}{\text{Inf}}\,[S^\rho(A) + S^\rho(B)], \tag{15}$$

is the optimum lower bound. So far, it has not been possible to obtain a constructive expression for $\Lambda_D(A, B)$ in the general case; but sharper and sharper lower bounds have been obtained and of course all of them imply that $\Lambda_D(A, B)$ vanishes essentially only when $A$, $B$ have a common eigenvector.

The first lower bound on $\Lambda_D(A, B)$ was obtained nearly twenty years ago by Deutsch [8] for the case when $A$, $B$ have non-degenerate spectra. Then (3a), (3b) reduce to

$$A = \sum_i a_i\,|a_i\rangle\langle a_i| \tag{16a}$$

$$B = \sum_j b_j\,|b_j\rangle\langle b_j|\,. \tag{16b}$$

For such observables, Deutsch showed that

$$\Lambda_D(A, B) \geq 2 \log\,[\,2\,/\,(1 + \underset{i,j}{\text{Sup}}\,|\langle a_i|b_j\rangle|\,)\,]\,. \tag{17}$$

Clearly the RHS in (17) is nonnegative and vanishes only when

$$\underset{i,j}{\text{Sup}}\,|\langle a_i|b_j\rangle| = 1, \tag{18}$$

which happens only when $A$, $B$ have a common eigenstate or have eigenstates arbitrarily close to each other. Deutsch inequality (17) was soon generalised to the case when the spectra of $A$, $B$ have degeneracies also, by Partovi [15], who showed that

$$\Lambda_D(A, B) \geq 2 \log\,[\,2\,/\,\underset{i,j}{\text{Sup}}\,\|P^A(a_i) + P^B(b_j)\|\,], \tag{19}$$

where the symbol ∥ ∥ stands for the operator norm. Partovi's inequality (19) reduces to Deutsch's inequality (17) when $A, B$ have non-degenerate spectra.

It was Kraus [16], who pointed out that the inequalities (18) and (20) are not optimal. Based on explicit calculations for 2, 3 and 4-dimensional examples, he conjectured that a much stronger inequality holds in finite-dimensions for the case of so-called "complementary observables". Two observables, $A, B$ with totally non-degenerate spectra are said to be complementary, if their eigenvectors satisfy

$$|<a_i|b_j>| = 1/\sqrt{n}, \qquad (20)$$

for all $1 \leq i, j \leq n$, where $n$ is the dimension of the Hilbert Space. Kraus' conjecture was that such complementary observables obey the optimal uncertainty relation

$$S^\rho(A) + S^\rho(B) \geq \Lambda_D(A, B) = \log(n). \qquad (21)$$

Kraus' conjecture was proved by Maassen and Uffink [17], who obtained the following lower bound on the sum of entropic uncertainties for any two observables with non-degenerate spectra in a finite-dimensional Hilbert Space:

$$\Lambda_D(A, B) \geq \log [\, 1 / \underset{i,j}{\text{Max}}\, |<a_i|b_j>|^2\,]. \qquad (22)$$

For the case of complementary observables, which satisfy (20), Maassen-Uffink bound (22) reduces to the form (21) conjectured by Kraus.

Recently Krishna and Parthasarathy [18] have generalised the result of Maassen and Uffink to obtain the following result, which is valid for any pair of observables in a finite-dimensional Hilbert Space:

$$\Lambda_D(A, B) \geq \log [\, 1 / \underset{i,j}{\text{Max}}\, \| P^A(a_i)P^B(b_j) \|^2\,], \qquad (23)$$

Clearly, Krishna-Parthasarathy bound (23) reduces to the Maassen-Uffink (and Kraus) bound (22) when the observables $A, B$ have non-degenerate spectra. Following the remarks made by Maassen and Uffink [17], it seems possible to extend the inequalities (22), (23) (by replacing Max with Sup), so as to be applicable also to the case when the Hilbert Space is infinite dimensional, as long as we restrict ourselves to observables with purely discrete spectra. Also, since

$$\| P^A(a_i)P^B(b_j) \|^2 = \| P^A(a_i)P^B(b_j)P^A(a_i) \| \leq \tfrac{1}{4} \| P^A(a_i) + P^B(b_j) \|^2, \qquad (24)$$

the Krishna-Parthasarathy bound (23) is stronger than the Partovi bound (19) and is thus the sharpest available bound on the sum of entropic uncertainties of two observables in the case of distinct measurements.

However, Garrett and Gull [19] showed in 1990 that the Maassen–Uffink bound (22) is not optimal in the case of a two-dimensional Hilbert Space when the two observables $A, B$, are not commuting or complementary. Recently Sanchez-Ruiz [20] has shown that the bound obtained by Garrett and Gull is indeed the optimal entropic uncertainty bound in two-dimensional Hilbert Space for distinct measurements. We shall consider this bound in section 4.

## 3. Optimal Entropic Uncertainty Relation for Successive Measurements

All the uncertainty relations discussed in section 2, refer to a situation where the observables $A$, $B$ are measured on two distinct but identically prepared ensembles of systems in state $\rho$. We shall now consider a situation where the observables $A$, $B$ are measured sequentially on the same ensemble of systems in state $\rho$. Then the joint probability $\Pr^\rho_{A,B}(a_i, b_j)$ that the outcomes are $a_i$, $b_j$ respectively, is given by the Wigner formula (4); and the probabilities $\Pr^\rho_{A,B}(a_i)$, $\Pr^\rho_{A,B}(b_j)$ for obtaining different outcomes in the $A$ and $B$ measurements, are the marginals of the above joint probability and are given by

$$\Pr^\rho_{A,B}(a_i) = \mathrm{Tr}\,[\rho P^A(a_i)] \tag{25a}$$

$$\Pr^\rho_{A,B}(b_j) = \mathrm{Tr}\,[\sum_i (P^A(a_i)\rho P^A(a_i))\,P^B(b_j)] = \mathrm{Tr}\,[\mathcal{E}(\rho)P^B(b_j)], \tag{25b}$$

where, as in (6), $\mathcal{E}(\rho)$ is given by

$$\mathcal{E}(\rho) = \sum_i P^A(a_i)\,\rho\,P^A(a_i). \tag{26}$$

Here, we may note the important feature of quantum theory that, unless $\mathcal{E}(\rho) = \rho$, we have

$$\Pr^\rho_{A,B}(b_j) \neq \mathrm{Tr}\,[\rho P^B(b_j)]. \tag{27}$$

Eq. (27) shows that the probability (25b) that the outcome $b_j$ is realised in a $B$-measurement, when an ensemble of systems prepared in state $\rho$ is subjected to the sequence of measurements $A$, $B$, is *different* from the probability $\mathrm{Tr}\,[\rho P^B(b_j)]$ that the same outcome is realised when there is no intervening measurement of observable $A$ prior to the measurement of $B$. This important feature of quantum theory has been referred to as the *quantum interference of probabilities* by de Broglie [21,6] and clearly expresses how a prior measurement influences the probability distributions of the outcomes of later measurements.

The uncertainties in the outcomes of $A$ and $B$, in a sequential measurement, are given by the information-theoretic entropies associated with the probability distributions (25a) and (25b)

$$S^\rho_{A,B}(A) = -\sum_i \Pr^\rho_{A,B}(a_i) \log \Pr^\rho_{A,B}(a_i) \tag{28a}$$

$$S^\rho_{A,B}(B) = -\sum_j \Pr^\rho_{A,B}(b_j) \log \Pr^\rho_{A,B}(b_j). \tag{28b}$$

By using the standard properties of the eigenprojectors of the observable $A$

$$P^A(a_i)\,P^A(a_j) = \delta_{ij}\,P^A(a_i) \tag{29a}$$

$$\sum_i P^A(a_i) = I, \tag{29b}$$

we can show the following:

$$S^\rho_{A,B}(A) = S^\rho(A) = S^{\mathcal{E}(\rho)}(A) \tag{30a}$$

$$S^\rho_{A,B}(B) = S^{\mathcal{E}(\rho)}(B), \tag{30b}$$

where $\mathcal{E}(\rho)$ is given by (26). From (30b) it follows that the entropies $S^\rho_{A,B}(B)$ and $S^\rho(B)$ are in general different unless $\mathcal{E}(\rho) = \rho$, which happens only when $\rho$ is a mixture of eigenstates of $A$.

The optimal uncertainty relation for successive measurements can be expressed in the form

$$S^\rho_{A,B}(A) + S^\rho_{A,B}(B) \geq \Lambda_S(A, B), \tag{31}$$

where the optimum bound is given by

$$\Lambda_S(A, B) = \underset{\rho}{\mathrm{Inf}} \, [S^\rho_{A,B}(A) + S^\rho_{A,B}(B)]. \tag{32}$$

From (30a) and (30b), we obtain

$$\Lambda_S(A, B) = \underset{\mathcal{E}(\rho)}{\mathrm{Inf}} \, [S^{\mathcal{E}(\rho)}(A) + S^{\mathcal{E}(\rho)}(B)]. \tag{33}$$

The map, $\rho \to \mathcal{E}(\rho)$, given by (26) maps the class of all density operators into a proper subset of itself, namely the set of all those density operators which commute with all the eigenprojectors $\{P^A(a_i)\}$ of $A$ (see for instance [6, chapter 7]). These are precisely those density operators, which are expressible as mixtures of eigen-states of $A$. Thus the infimum on the RHS of (33) is over a much smaller class of density operators (namely those of the form $\mathcal{E}(\rho)$) than the set of all density operators which occurs in the RHS of (15). Thus, we are led to the inequality

$$\Lambda_S(A, B) \geq \Lambda_D(A, B), \tag{34}$$

which shows that the optimal bound on the sum of uncertainties for successive measurements of two observables is always greater than or equal to the optimal bound for sum of uncertainties when the same observables are measured on distinct ensemble of systems. While there is no general constructive expression for the latter, we shall show that the optimal bound $\Lambda_S(A, B)$ can be expressed in terms of the eigenstates of the observables $A$, $B$, when they have non-degenerate spectra.

A lower bound on $\Lambda_S(A, B)$ was obtained some time ago [10] by making use of the joint entropy $S^\rho_{A,B}(A, B)$, defined in terms of the joint probabilities (4):

$$S^\rho_{A,B}(A, B) = - \sum_{i,j} \mathrm{Pr}^\rho_{A,B}(a_i, b_j) \, \log \mathrm{Pr}^\rho_{A,B}(a_i, b_j). \tag{35}$$

Since the probabilities $\Pr^\rho_{A,B}(a_i)$ and $\Pr^\rho_{A,B}(b_j)$, given by (25a), (25b), are marginals of the joint probability (4), we have the sub-additivity inequality:

$$S^\rho_{A,B}(A) + S^\rho_{A,B}(B) \geq S^\rho_{A,B}(A, B). \tag{36}$$

Incidentally, we also have the inequalities,

$$S^\rho_{A,B}(A, B) \geq S^\rho_{A,B}(A) = S^\rho(A) \tag{37a}$$

$$S^\rho_{A,B}(A, B) \geq S^\rho_{A,B}(B), \tag{37b}$$

which, along with (36), can be used to define conditional entropies and mutual information, for sequential measurements of two observables in quantum theory.

Now, from the equations (35) and (4) which define the joint entropy and the joint probabilities, we can easily obtain the inequality

$$S^\rho_{A,B}(A, B) \geq \log [1 / \sup_{i,j} \| P^A(a_i) P^B(b_j) P^A(a_i) \|]. \tag{38}$$

From (36) and (38), we obtain the uncertainty relation

$$S^\rho_{A,B}(A) + S^\rho_{A,B}(B) \geq \log [1 / \sup_{i,j} \| P^A(a_i) P^B(b_j) P^A(a_i) \|], \tag{39}$$

where the bound on the RHS is the same as the Krishna-Parthasarathy bound on the sum of uncertainties for distinct measurements.

We can obtain a much stronger bound than (39) for the sum of uncertainties for successive measurements. Our result on the optimal uncertainty bound for successive measurements is contained in the following Theorem 1.

**Theorem 1:** The optimal bound $\Lambda_S(A, B)$ on the sum of entropic uncertainties of two observables $A, B$, with purely discrete spectra, when they are measured sequentially on the same ensemble of systems, is given by

$$\Lambda_S(A, B) = \inf_i \inf_{P^A(a_i)|\psi\rangle = |\psi\rangle} - \sum_j \langle\psi|P^B(b_j)|\psi\rangle \log \langle\psi|P^B(b_j)|\psi\rangle, \tag{40a}$$

where the states $|\psi\rangle$ are assumed to be normalized. When the observable $A$ has non-degenerate spectrum, eq (40a) reduces to

$$\Lambda_S(A, B) = \inf_i - \sum_j \langle a_i|P^B(b_j)|a_i\rangle \log \langle a_i|P^B(b_j)|a_i\rangle. \tag{40b}$$

When both $A$ and $B$ have non-degenerate spectra, the optimal lower bound further reduces to

$$\Lambda_S(A, B) = \inf_i - \sum_j |\langle a_i|b_j\rangle|^2 \log |\langle a_i|b_j\rangle|^2. \tag{40c}$$

**Proof:** We start with the expression for the optimum bound $\Lambda_S(A, B)$, given by (33), where the infimum is to be taken only over states of the form $\mathcal{E}(\rho)$, given by (26). Since the entropies $S^{\mathcal{E}(\rho)}(A)$ and $S^{\mathcal{E}(\rho)}(B)$, are concave functions of $\rho$, we need to take the infimum in (33) only over pure states which occur in the decomposition of $\mathcal{E}(\rho)$. Since, as we have already noted, states of the form $\mathcal{E}(\rho)$ are mixtures of eigenstates of $A$, we need to take infimum in (33) only over the eigenstates of $A$. In each eigenstate $|\psi\rangle$ of $A$, the entropy $S^{|\psi\rangle}(A)$ vanishes and therefore eq. (33) reduces to

$$\Lambda_S(A, B) = \underset{i}{\text{Inf}} \underset{P^A(a_i)|\psi\rangle = |\psi\rangle}{\text{Inf}} S^{|\psi\rangle}{}_{A,B}(B). \tag{41}$$

Now, if we employ eq. (28b) for the entropy $S^{|\psi\rangle}{}_{A,B}(B)$ in eq. (41), we are immediately led to eq. (40a) for the optimal bound. Eqs. (40b) and (40c) are direct consequences of eq. (40a) when the spectra of $A$ and $B$ turn out to be non-degenerate. This completes the proof of Theorem 1.

From eq.(40c) it follows that for observables with non-degenerate spectra

$$\Lambda_S(A, B) \geq \log [1 / \underset{i,j}{\text{Sup}} |\langle a_i|b_j\rangle|^2]. \tag{42}$$

The fact that the optimal uncertainty bound for successive measurements is greater than the Maassen-Uffink bound for distinct measurements could have anyway been inferred from eqs. (34) and (22).

The bound on the RHS of eq. (40c), for the sum of uncertainties in successive measurements, has also been obtained in a recent investigation by Cerf and Adami [22]. They have also noted the important fact that this bound is greater than the Maassen-Uffink bound for distinct measurements. Our derivation above shows that the RHS of (40c) actually gives the optimal bound on the sum of uncertainties in the outcomes of successive measurements of observables with non-degenerate spectra.

From the above derivation it also follows that $S^\rho{}_{A,B}(A) + S^\rho{}_{A,B}(B)$ attains its infimum in eigenstates of observable $A$, states in which $S^\rho{}_{A,B}(A)$ vanishes. Therefore, the optimal lower bound of $S^\rho{}_{A,B}(A) + S^\rho{}_{A,B}(B)$ is also the optimal lower bound of $S^\rho{}_{A,B}(B)$, or equivalently,

$$S^\rho{}_{A,B}(B) \geq \Lambda_S(A, B). \tag{43}$$

In particular, when $A$, $B$ have non-degenerate spectra,

$$S^\rho{}_{A,B}(B) \geq \underset{i}{\text{Inf}} - \underset{j}{\sum} |\langle a_i|b_j\rangle|^2 \log |\langle a_i|b_j\rangle|^2. \tag{43a}$$

When the RHS of (43) is non-zero, which invariably happens when $A$, $B$ do not have a common eigenvector, the inequality (43) gives a non-trivial lower bound on the uncertainty in the outcome of $B$-measurement, which arises essentially because the ensemble of systems prepared in state $\rho$, has been first subjected to an *A*-measurement. Had there been no such intervening measurement on the ensemble, then the uncertainty $S^\rho(B)$ in the outcome of *B*-measurement can be made arbitrarily small, in fact zero by choosing the initial state $\rho$ to be a mixture of eigenstates of *B*. Equation (43) shows that whenever *A,B* do not have a joint eigenstate, the outcome of a *B*-measurement which follows an *A*-measurement is *always uncertain whatever be the initial state of the system*, and the optimal lower bound on this uncertainty is again given by the RHS of (40a)-(40c).

Using the method outlined above, we can obtain optimal bounds on the sum of entropic uncertainties of any arbitrary sequence of measurements, provided we restrict ourselves to observables with discrete spectra. For instance, if *C* is another observable with spectral decomposition

$$C = \sum_k c_k P^C(c_k), \qquad (44)$$

then the joint probability $\text{Pr}^\rho_{A,B,C}(a_i, b_j, c_k)$, that the outcomes $a_i, b_j, c_k$, are realised in a sequential measurement of *A, B, C*, is given by the Wigner formula:

$$\text{Pr}^\rho_{A,B,C}(a_i, b_j, c_k) = \text{Tr}\,[P^C(c_k)P^B(b_j)P^A(a_i)\,\rho\,P^A(a_i)P^B(b_j)P^C(c_k)]\,. \qquad (45)$$

From (45), we obtain the joint entropy

$$S^\rho_{A,B,C}(A, B, C) = -\sum_{i,j,k} \text{Pr}^\rho_{A,B,C}(a_i, b_j, c_k)\,\log \text{Pr}^\rho_{A,B,C}(a_i, b_j, c_k). \qquad (46)$$

The uncertainty in the outcomes of the *C*-measurement, in a sequential measurement of *A, B, C*, is given by

$$S^\rho_{A,B,C}(C) = -\sum_k \text{Pr}^\rho_{A,B,C}(c_k)\,\log \text{Pr}^\rho_{A,B,C}(c_k). \qquad (47)$$

Using the standard properties of spectral projectors of *A* and *B*, we can show that

$$S^\rho_{A,B,C}(C) = S^{\Gamma\mathcal{E}(\rho)}(C), \qquad (48)$$

Where, $\mathcal{E}$ is as given by eq. (26) and $\Gamma$ is given by

$$\Gamma(\rho) = \sum_j P^B(b_j)\,\rho\,P^B(b_j) \qquad (49)$$

From the fact that the probability distributions of the observables *A, B, C*, are marginals of the joint probability $\text{Pr}^\rho_{A,B,C}(a_i, b_j, c_k)$ given by (45), we can deduce the sub-additivity and the strong sub-additivity properties

$$S^\rho_{A,B,C}(A) + S^\rho_{A,B,C}(B) + S^\rho_{A,B,C}(C) \geq S^\rho_{A,B,C}(A, B, C), \qquad (50)$$

$$S^\rho_{A,B,C}(A, B) + S^\rho_{A,B,C}(B, C) \geq S^\rho_{A,B,C}(A, B, C) + S^\rho_{A,B,C}(B), \qquad (51)$$

in the same manner as in classical information theory.

The optimal entropic uncertainty relation for the successive measurement of three observables $A$, $B$, $C$, is given by

$$S^\rho_{A,B,C}(A) + S^\rho_{A,B,C}(B) + S^\rho_{A,B,C}(C) \geq \Lambda_S(A, B, C), \qquad (52)$$

where

$$\Lambda_S(A, B, C) = \inf_\rho [S^\rho_{A,B,C}(A) + S^\rho_{A,B,C}(B) + S^\rho_{A,B,C}(C)]. \qquad (53)$$

From (30a), (30b) and (48), we obtain

$$\Lambda_S(A, B, C) = \inf_{\mathcal{E}(\rho)} [S^{\mathcal{E}(\rho)}(A) + S^{\mathcal{E}(\rho)}(B) + S^{\Gamma\mathcal{E}(\rho)}(C)]. \qquad (54)$$

Again, in the RHS of eq. (54), we need to take infimum only over states of the form $\mathcal{E}(\rho)$, and following the line of argument in the proof of Theorem 1, we can easily establish the following Theorem 2, where for the sake of simplicity we give only the result for observables with non-degenerate spectra.

**Theorem 2:** The optimal bound $\Lambda_S(A,B,C)$ on the sum of entropic uncertainties of three observables $A$, $B$, $C$, with purely discrete and non-degenerate spectra, which are sequentially measured on the same ensemble of systems, is given by

$$\Lambda_S(A, B, C) = \inf_i - \sum_j |\langle a_i|b_j\rangle|^2 \log |\langle a_i|b_j\rangle|^2$$

$$+ \inf_i - \sum_k \{ \sum_j (|\langle a_i|b_j\rangle|^2 |\langle b_j|c_k\rangle|^2) \log [\sum_j |\langle a_i|b_j\rangle|^2 |\langle b_j|c_k\rangle|^2] \}. \qquad (55)$$

There are two terms in the optimal bound (55). The first is nothing but the lower bound $\Lambda_S(A, B)$ that was derived earlier in eq. (40c), for the sequential measurement of the two observables, $A$, $B$; it is also the optimal lower bound (43a) on the uncertainty in the outcomes of the B-measurement, $S^\rho_{A,B}(B) = S^\rho_{A,B,C}(B)$ in a sequential measurement of $A$, $B$, $C$. The second term (55) can be easily shown to be the optimal bound on the uncertainty in the outcomes of $C$-measurement when a sequential measurement of $A$, $B$, $C$ is carried out. In other words, we have

$$S^\rho_{A,B,C}(C) \geq \inf_i - \sum_k \{ \sum_j (|\langle a_i|b_j\rangle|^2 |\langle b_j|c_k\rangle|^2) \log [\sum_j |\langle a_i|b_j\rangle|^2 |\langle b_j|c_k\rangle|^2] \}. \qquad (56)$$

Since the matrix, $U_{jk} = |\langle b_j|c_k\rangle|^2$, is a doubly stochastic matrix, it follows [9] that the lower bound on $S^\rho_{A,B,C}(C)$ given by the RHS of eq. (56) is greater than or equal to the lower bound on $S^\rho_{A,B}(B)$ given by the RHS of eq. (43a). The above results (55), (56) can be trivially extended to arbitrary sequence of measurements of observables with purely discrete and non-degenerate spectra.

**4. Optimal Uncertainty Relations in Two-dimensional Hilbert Space.**

We have seen above in eq. (34) that the optimal bound $\Lambda_S(A, B)$ on the sum of uncertainties of observables A, B, when they are measured sequentially on the same ensemble of systems, is greater than or equal to the optimal bound $\Lambda_D(A, B)$ when they are measured on distinct but identically prepared ensemble of systems. However, the question still remains as to whether $\Lambda_S(A, B)$ is strictly greater than $\Lambda_D(A, B)$ for at least some pair of observables $A, B$. In order to decide this issue we need to consider a situation for which the optimal bound $\Lambda_D(A, B)$ has been determined for a large enough class of observables.

Recently Sanchez-Ruiz [20] has shown that the uncertainty bounds derived sometime ago by Garrett and Gull [19] for observables in a two-dimensional Hilbert Space are in fact optimal. We shall utilise this bound to show that $\Lambda_S(A, B)$ is indeed strictly greater than $\Lambda_D(A, B)$ for a large class of observables.

Let $A, B$ refer to the components of spin, for a spin-½ system, along the directions $n_1, n_2$, i.e. $A = \sigma.n_1$ and $B = \sigma.n_2$. If $n_1.n_2 = \cos\theta$, we have

$$|\langle a_1|b_1\rangle|^2 = |\langle a_2|b_2\rangle|^2 = \cos^2\theta/2 \qquad (57a)$$

$$|\langle a_1|b_2\rangle|^2 = |\langle a_2|b_1\rangle|^2 = \sin^2\theta/2 \qquad (57b)$$

The optimal bound (40c), on the sum of uncertainties of $\sigma.n_1$ and $\sigma.n_2$, when they are measured sequentially on the same ensemble of systems, becomes

$$\Lambda_S(\sigma.n_1, \sigma.n_2) = \Lambda_S(\theta) = -\cos^2\theta/2 \log \cos^2\theta/2 - \sin^2\theta/2 \log \sin^2\theta/2, \qquad (58)$$

which is nothing but the entropic uncertainty of $\sigma.n_2$ in the eigenstates of $\sigma.n_1$.

When $\sigma.n_1$ and $\sigma.n_2$ are measured on distinct but identically prepared ensembles of systems, the uncertainty bound (17) obtained by Deutsch becomes

$$\Lambda_D(\sigma.n_1, \sigma.n_2) \geq \Lambda_{D1}(\theta) = 2\log(2/[1+\text{Max}\{|\cos\theta/2|, |\sin\theta/2|\}]) \qquad (59)$$

The stronger bound (22) obtained by Maassen and Uffink becomes

$$\Lambda_D(\sigma.n_1, \sigma.n_2) \geq \Lambda_{D2}(\theta) = 2 \log (1 / \text{Max} \{\cos^2\theta/2, \sin^2\theta/2\}). \qquad (60)$$

Recently, Sanchez-Ruiz has shown that the optimum bound on the sum of uncertainties of $\sigma.n_1$ and $\sigma.n_2$, when they are measured on distinct but identically prepared ensembles of systems, is given by the following:

$$\Lambda_D(\sigma.n_1, \sigma.n_2) = \Lambda_D(\theta) = -2 \cos^2\theta/4 \log \cos^2\theta/4$$

$$- 2 \sin^2\theta/4 \log \sin^2\theta/4, \qquad (61)$$

when $0 \leq \theta \leq \theta^*$, where $\theta^*$ is given by the transcendental equation

$$\cos(\theta^*/2) \log [(1+\cos\theta^*/2)/(1-\cos\theta^*/2)] = 2, \qquad (62)$$

which implies $\theta^* \approx 67°$. Eq. (61) shows that the sum of uncertainties in $\sigma.n_1$ and $\sigma.n_2$, takes on the infimum value in the eigenstates of $\sigma.(n_1 + n_2)$ when $0 \leq \theta \leq \theta^*$. When $\pi - \theta^* \leq \theta \leq \pi$, the optimum bound is

$$\Lambda_D(\sigma.n_1, \sigma.n_2) = \Lambda_D(\theta) = -\cos^2(\pi/4 + \theta/4) \log \cos^2(\pi/4 + \theta/4)$$

$$-\sin^2(\pi/4 + \theta/4) \log \sin^2(\pi/4 + \theta/4)$$

$$-\cos^2(\pi/4 - \theta/4) \log \cos^2(\pi/4 - \theta/4)$$

$$-\sin^2(\pi/4 - \theta/4) \log \sin^2(\pi/4 - \theta/4). \quad (63)$$

In this case, the optimum bound is attained in the eigenstates of $\sigma.(n_1 - n_2)$. When $\theta^* \leq \theta \leq \pi - \theta^*$, the optimal bound $\Lambda_D(\theta)$ seems to have no analytical expression and has to be evaluated numerically.

The numerical values of $\Lambda_{D1}(\theta)$ (the Deutsch bound), $\Lambda_{D2}(\theta)$ (the Maassen-Uffink bound) and the optimum bound $\Lambda_D(\theta)$, for the case of distinct measurements, have been tabulated by Garrett and Gull [19]. In Table 1 we give these values along with the numerical values of the optimum bound $\Lambda_S(\theta)$ for successive measurements given by eq. (58).

**Table 1.** Numerical Estimates of Different Uncertainty Bounds

| $\theta°$ | $\Lambda_S(\theta)$ | $\Lambda_D(\theta)$ | $\Lambda_{D2}(\theta)$ | $\Lambda_{D1}(\theta)$ |
|---|---|---|---|---|
| 0 | 0.000 | 0.000 | 0.000 | 0.000 |
| 10 | 0.045 | 0.028 | 0.008 | 0.004 |
| 20 | 0.135 | 0.089 | 0.031 | 0.015 |
| 30 | 0.246 | 0.173 | 0.069 | 0.034 |
| 40 | 0.361 | 0.271 | 0.124 | 0.061 |
| 50 | 0.469 | 0.378 | 0.197 | 0.096 |
| 60 | 0.562 | 0.492 | 0.288 | 0.139 |
| 70 | 0.633 | 0.604 | 0.399 | 0.190 |
| 80 | 0.678 | 0.673 | 0.533 | 0.249 |
| 90 | 0.693 | 0.693 | 0.693 | 0.317 |

Note: All logarithms are calculated to base $e$.

From Table 1, we see that

$$\Lambda_S(\theta) \geq \Lambda_D(\theta) \geq \Lambda_{D2}(\theta) \geq 2\Lambda_{D1}(\theta), \quad (64)$$

where, equality holds only when $\theta = 0, \pi/2$. Hence, in a two-dimensional Hilbert Space, the optimal bound $\Lambda_S(\sigma.n_1, \sigma.n_2)$ for successive measurements of $\sigma.n_1$ and $\sigma.n_2$, is *strictly greater* than the optimal bound $\Lambda_D(\sigma.n_1, \sigma.n_2)$ when the observables $\sigma.n_1, \sigma.n_2$ are measured on distinct ensembles, except when $n_1, n_2$ are parallel or perpendicular, i.e. when the observables $\sigma.n_1, \sigma.n_2$ either commute or are complementary.

## 5. Conclusion

We have seen that there are actually two formulations of the uncertainty principle in quantum theory: one for the case of *distinct measurements* performed on two *distinct but identically prepared ensembles* of systems; and another for the case of *successive measurements* performed on the on the *same ensemble* of systems. Consequently there are two classes of uncertainty relations, which correspond to these two different experimental situations.

The standard variance form of the uncertainty relation due to Robertson, or the entropic form uncertainty relations due to Everett and Deutsch and improved upon by later researchers, all deal with measurements performed on distinct but identically prepared ensembles of systems. It is now well understood that these uncertainty relations essentially reflect the limitations imposed by quantum theory on the preparation of ensembles of systems in identical states. They do not in any way express the influence of measurement of one observable on the uncertainty in the outcomes of another, contrary to what seems to have been envisaged by Heisenberg in his pioneering work on the uncertainty principle.

On the other hand, a different class of uncertainty relations can be formulated for the case when a set of observables is sequentially measured on the same ensemble of systems. These uncertainty relations clearly reveal the influence of measurement of one observable on the uncertainties in the outcomes of later measurements. The fact that the optimal lower bound on the sum of uncertainties of successive measurements turns out to be in general not less than and often strictly greater than the optimal bound on the corresponding sum of uncertainties for measurements performed on distinct ensembles of systems, shows that in quantum theory measurements do have a profound influence on the uncertainties in the outcomes of all subsequent observations.

As regards the particular physical situation discussed by Heisenberg in his pioneering paper, viz. the measurement of position followed by that of momentum or vice versa, this has remained outside the purview of the standard mathematical formulation of quantum theory because of the difficulties associated with the extension of the conventional collapse postulate for observables such as position and momentum, which have continuous spectra. Various investigations [6, chapters 7,8] have however shown that any meaningful generalization of the collapse postulate would necessarily imply that the state of a system immediately after a precise measurement of position would be such that its momentum distribution is entirely concentrated at $\pm \infty$, or vice versa. This is perhaps the extent to which Heisenberg's intuitive understanding of the quantum uncertainty principle can be vindicated on the basis of the mathematical formalism of quantum mechanics.